\documentclass[10pt, prd, twocolumn, showpacs, amsmath, amssymb, aps, floats, floatfix, nofootinbib]{revtex4-1}

\usepackage{bm}
\usepackage{array}
\usepackage{graphicx}
\usepackage{subfigure}
\usepackage{multirow}
\usepackage{dcolumn}
\usepackage[]{color}
\usepackage[CJKbookmarks=true,colorlinks,linkcolor=blue,anchorcolor=red,citecolor=green]{hyperref}
\usepackage{cleveref}
\usepackage{multirow}
\usepackage{setspace}

\def\kpc{\mathrm{kpc}}
\def\km{\mathrm{km}}

\def\GeV{\mathrm{GeV}}

\def\eV{\mathrm{eV}}

\def\cm{\mathrm{cm}}
\def\s{\mathrm{s}}

\newcolumntype{p}{D{,}{\pm}{-1}}

\begin{document}
\title{Impact of a nearby subhalo on the constraint of dark matter annihilation from cosmic ray antiprotons
}

\author{Yi Zhao$^{1}$}
\author{Xiao-Jun Bi$^{2,3}$}
\author{Su-Jie Lin$^{4}$}
\author{Peng-Fei Yin$^{2}$}

\affiliation{
$^1$ College of Physics and Materials Science, Tianjin Normal University, Tianjin 300387, China \\
$^2$ Key Laboratory of Particle Astrophysics, Institute of High Energy Physics, Chinese Academy of Sciences, Beijing 100049, China \\
$^3$ School of Physical Sciences, University of Chinese Academy of Sciences, Beijing 100049, China \\
$^4$ Zhuhai Campus of Sun Yat-Sen University, Sun Yat-Sen University, Zhuhai, Guangdong, 519082, China
}

\begin{abstract}
Numerous simulations indicate that a large number of subhalos should be hosted by the Milky Way.
The potential existence of a nearby subhalo could have important implications for our understanding of dark matter (DM) annihilation.
In this study, we investigate the hypothetical presence of a nearby subhalo and set the upper limits on the DM annihilation cross section by analyzing the cosmic-ray antiproton spectrum.
By presenting the ratios of annihilation cross section limits for scenarios with and without a nearby subhalo,
we can quantitatively evaluate the potential impact of the nearby subhalo on the limits of the DM annihilation cross section.
The impacts of the concentration model and the subhalo probability distribution have been considered.
We explore the antiproton contribution of the potential nearby DM subhalo accounting for the
DAMPE $e^\pm$ spectrum at $\sim 1.4$ TeV and find that the current AMS-02 antiproton results do not place the constraint on this contribution.

\end{abstract}

%\date{\today}

\maketitle

\section{Introduction}

The nature of dark matter (DM) remains one of the most pressing questions in modern astrophysics, cosmology, and particle physics.
In the standard cosmology scenario, the formation of luminous galaxies
takes place within the halos of cold and collisionless non-baryonic DM.
Traditionally, the DM distribution has been modeled as a smoothly distributed component with a spherically symmetric
density profile, such as an isothermal or Navarro-Frenk-White (NFW) profile~\cite{Navarro:1996gj}.
However, a large number of theoretical arguments suggest that the DM distribution in
galactic halos may exhibit significant substructures~\cite{Klypin:1997fb,Ghigna:1998vn,Klypin:1999uc,Moore:1999nt,Boehm:2000gq,Green:2003un,Green:2005fa,Green:2005kf,
Bertschinger:2006nq,Berezinsky:2007qu,Diemand:2007qr,Loeb:2005pm,Diemand:2005vz,Gao:2005hn,Diemand:2006ey,Springel:2000qu,Zentner:2003yd}.

DM halos are thought to form through gravitational amplification of initial fluctuations in a bottom-up manner,
in which smaller objects collapse and merge into larger ones over time.
Extensive N-body simulations have revealed that some of the total halo mass can endure tidal disruption, appearing as distinct subhalos within host halos.~\cite{Tormen:1997ik,Klypin:1999uc,Ghigna:1999sn,Wilson:2012fx,DeLucia:2003xe,Kravtsov:2004cm}.
The existence of subhalos in galactic halos has been confirmed through numerous high-resolution numerical simulations.
This is a key aspect of the standard cosmology with hierarchical structure formation.

Simulations indicate that the Milky Way should host hundreds of subhalos~\cite{Klypin:1999uc,Kauffmann:1993gv,Gao:2004au,Springel:2008cc,Diemand:2008in,Garrison-Kimmel:2013eoa}.
However, astrophysical mechanisms suggest that many subhalos are challenging to observe as they are non-luminous and not very massive.
The number of satellite galaxies observed around the Milky Way is smaller
than the projected number of DM subhaloes as predicted by the cold DM model.
This suggests the existence of a substantial
quantity of exceedingly dim galaxies and DM-dominated haloes with minimal or negligible
stellar content within the Local Group.

If the DM particles exist in the form of weakly interacting particles, which are well-motivated
DM candidates, subhalos can be traced through the products resulting from DM annihilation.
They can boost the fluxes of annihilation products, potentially enabling their detection in cosmic ray observations.
Such observations not only offer a unique avenue for exploring the particle properties of DM, but also could  provide insights into the DM structure.

The study of antimatter particles is particularly significant in this field.
The accurate quantification of antiproton spectra by space-borne instruments like AMS-02 offers excellent sensitivity to probe DM particles.
These antiprotons are expected to primarily originate from inelastic collisions
between cosmic rays and the interstellar medium. However, they may also
be generated through DM annihilation or decay.
For example, the tentative excess at ~$\mathcal{O}(10)$ GeV of the AMS-02 antiproton data can be explained by DM annihilation~\cite{Cuoco:2016eej,Cui:2016ppb}. Various discussions on this issue can be found in the literature ~\cite{Cui:2018klo,Cuoco:2019kuu,Cholis:2019ejx,Lin:2019ljc,Boudaud:2019efq,Heisig:2020nse,Kahlhoefer:2021sha,Zhu:2022tpr,Lv:2023gdt}.

In this study, we specifically examine the impact of the possible nearby subhalo on antiprotons induced by DM annihilation. The
constraints on the thermally averaged annihilation cross section of DM $\langle\sigma v\rangle$ are investigated by using the antiproton spectra
measured by AMS-02 ~\cite{AMS:2021nhj}. Subsequently, we propose the hypothesis that a DM subhalo of a certain mass and distance from the Earth exists and investigate how
such a subhalo influences the limits on $\langle\sigma v\rangle$.
Meanwhile, we examine how the concentration model and potential subhalo
distribution could impact this study individually. We also explore the antiproton limits for the potential nearby DM subhalo accounting for the
DAMPE $e^\pm$ spectrum at $\sim 1.4$ TeV ~\cite{DAMPE:2017cev}.

This paper is organized as follows. In Sec. II, we briefly introduce the calculation of the antiprotons from the cosmic ray interactions and the smooth DM component. In Sec. III, we set the constraints on the DM $\langle\sigma v\rangle$ with the contribution from the nearby DM subhalo. The contribution from a series of DM subhalos is also discussed.
In Sec. IV, we investigate the antiproton limits for the potential nearby DM subhalo accounting for the DAMPE $e^\pm$ spectrum at $\sim 1.4$ TeV.
Finally, Sec. V is the conclusion.

\section{Antiprotons from cosmic ray interactions and smooth DM component}
\label{sec_2}

In this work, we utilize the numerical tool GALPROP~\cite{Moskalenko:1997gh,Strong:1998pw} to derive the contribution of antiprotons from the cosmic ray interactions and the smooth DM component.
The propagation of cosmic rays in the Milky Way can be characterized by the diffusive transport equation~\cite{Gaisser:2016uoy}, which includes the following primary propagation parameters: diffusion coefficient at the reference rigidity $D_0$, diffusion coefficient index $\delta$, Alfvenic speed $v_A$ for reacceleration purposes, half-height of the propagation halo $z_h$, assumed broken power-law in rigidity with indices $\nu_1$ and $\nu_2$ below or above the break rigidity $R_{br}$ for the injection spectrum of cosmic-ray nuclei, and propagated flux normalization of protons $A_p$.
As the DM contributions are involved, the assumptions of the DM mass, thermally averaged annihilation cross section, annihilation channel, and density profile are also required.
For the solar modulation effect, we consider the simple force field model with one parameter of the modulation potential $\Phi$.

We find that the diffusion-reacceleration propagation model is better than the diffusion-convection model when fitting the antiproton flux measured by AMS-02, as has been previously reported~\cite{Cui:2016ppb,Yuan:2017ozr}.
Therefore, we adopt the diffusion-reacceleration propagation model in this work. For the main propagation parameters, we take the values $D_0=7.24\times10^{28}\cm^{2} \cdot\s^{-1}$, $\delta=0.38$, $v_A=38.5\; \km\cdot\s^{-1}$, $z_h=5.93 \;\kpc$, $\nu_1=1.69$, $\nu_2=2.37$, $\log_{10}(R_{br})=4.11$, and $\log_{10}(A_p)=-8.347$, which are obtained from the Markov chain Monte Carlo fit in Ref.~\cite{Yuan:2017ozr}.
Afterward, we free the solar modulation $\Phi$ in the range of $(0.3, 0.8 )$ GV.
We also add an energy-independent rescaling factor $\kappa$ for the secondary antiproton flux in the fit, which is used to describe the uncertainty of the antiproton production cross sections in proton-proton collisions~\cite{diMauro:2014zea}.
This factor is allowed to vary within $(0.5,2.0)$ in the fit but is about equal to 1 for the best-fit results.
The DM density distribution is adopted to be the NFW profile~\cite{Navarro:1996gj}
\begin{eqnarray}\label{eq0}
\rho(r)=\frac{\rho_s}{(r/r_s)(1+r/r_s)^2}.
\end{eqnarray}
For the Galactic halo, we take $\rho_s=0.35\;\GeV\cdot\cm^{-3}$ and $r_s=20\;\kpc$, which
correspond to a local DM density of $0.4\;\GeV\cdot\cm^{-3}$ near the Solar system.

\begin{figure}[!htb]
\centering
\includegraphics[width=0.9\columnwidth, angle=0]{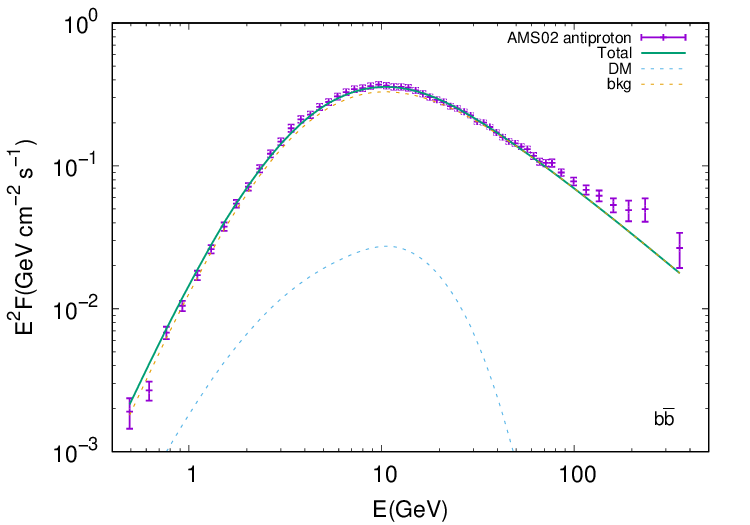}
\caption{The total antiproton flux (green solid line), comprised of
the secondary antiprotons (orange dashed line) and those generated by
DM annihilation (blue dashed line). The DM mass and $\langle\sigma v\rangle$ are 75 GeV and 0.9$\times \rm 10^{-26}cm^{3}s^{-1}$, respectively.}
\label{fig1}
\end{figure}

In Fig. \ref{fig1}, we take DM with a mass of 75 GeV as an example to show the antiproton prediction, comparing with the latest AMS-02 results ~\cite{AMS:2021nhj}. We find that
the experimental data can be well fitted by the antiprotons from cosmic-ray interactions, except that
the antiproton flux at ~$\mathcal{O}(10)$ GeV is potentially compensated by DM annihilation. Note that the significance of the antiproton excess at ~$\mathcal{O}(10)$ GeV and corresponding DM interpretation depends on many complicated factors, such as the secondary antiproton production process, propagation, solar modulation, and unclear experimental correlated errors ~\cite{Cui:2018klo,Cuoco:2019kuu,Cholis:2019ejx,Lin:2019ljc,Boudaud:2019efq,Heisig:2020nse,Kahlhoefer:2021sha,Zhu:2022tpr,Lv:2023gdt}. In this work, we focus on the impact of the nearby DM subhalo on the DM constraints and do not discuss this antiproton excess and the corresponding DM interpretation.

\section{DM annihilating to $\bar{p}$ with nearby subhalo}
\label{sec_2}

In this section, we consider the DM contribution to antiprotons
from the nearby subhalo, aiming to understand the extent of the impact of the nearby subhalo
on the constraint of the DM $\langle\sigma v\rangle$.
We take the nearby subhalo as an extended source and calculate its flux as
\begin{eqnarray}\label{eq1}
\phi^{r,E}=\int _{0}^{\pi} 2 \pi \sin \theta  d\theta \int _{0}^{\infty} \phi_{c}^{r,E}(r^\prime,\theta)r^{\prime2}dr^\prime ,
\end{eqnarray}
with
\begin{eqnarray}\label{eq3}
\phi_{c}^{r,E}({r^\prime},{\theta})={\rho}^2 \frac{dN}{dE} \frac{\langle\sigma v\rangle}{4m_{\chi}^2}C(d) ,
\end{eqnarray}
where $r^\prime$ is the distance from the halo center to the site of DM annihilation, $r$ is the distance from the subhalo center to
the Earth, $\theta$ is
the angle between $\vec{r}$ and $\vec{r}^\prime$, $\langle\sigma v\rangle$ is the thermally averaged DM annihilation cross section,
$dN/dE$ is the initial DM annihilation spectrum obtained by PPPC~\cite{Cirelli:2010xx,Ciafaloni:2010ti},
$m_\chi$ is the DM mass, and $\rho$ is the DM density.
The propagation term $C(d)$ is given by
\begin{eqnarray}\label{eq4}
C(d)=\int _{0}^{\infty} \frac{1}{(4 \pi D\tau)^{\frac{3}{2}}} e^{-\frac{d^2}{4D\tau}}d\tau,
\end{eqnarray}
where $D$ is the diffusion coefficient, $\tau$ is the delay time between the particle emission and observation, and $d=\sqrt{r^2+r^{\prime2}-2rr^\prime \cos\theta}$ is the distance from the Earth to the position where DM particles annihilate.

The density profile of subhalos within the solar neighborhood demonstrated a suitable correspondence with
the NFW profile, as evidenced by the results from high-resolution N-body simulations~\cite{Springel:2008cc}.
Consequently, we employ the NFW profile as a representation for the DM density within the subhalo
and determine the subhalo mass using the concentration model along with the parameters $r_s$ and $\rho_s$ in the NFW profile.

The concentration parameter is defined as the ratio of the virial radius to the scale radius of the halo $C=r_v/r_s$. This parameter describes the shape of the DM halo,
offering a nuanced understanding of how DM is distributed throughout the cosmic structure.
The concentration model can greatly influence the gravitational effects and interactions within galaxies and clusters.
It depends on the mass and formation history of the halo and is typically determined
from numerical simulations of structure formation.
Here we adopt the concentration model in Ref.~\cite{Yuan:2017ysv} (referred to as model 1), which is an approximation for the solar neighborhood based on the results of the N-body simulation Aquarius \cite{Springel:2008cc}. The concentration parameter $C$ is determined by the relation
\begin{eqnarray}\label{eq5}
8.66\times 10^6 (\frac{M_{\rm sub}}{10^6 M_{\odot}})^{-0.18}=\frac{200}{3}\frac{C^3}{\ln(1+C)-\frac{C}{(1+C)}},
\end{eqnarray}
where $M_{\rm sub}$ is the mass of subhalo and $M_{\rm \odot}$ is the solar mass.

\begin{figure*}[htb]
\centering
\subfigure[]{
  \includegraphics[width=1.0\columnwidth, angle=0]{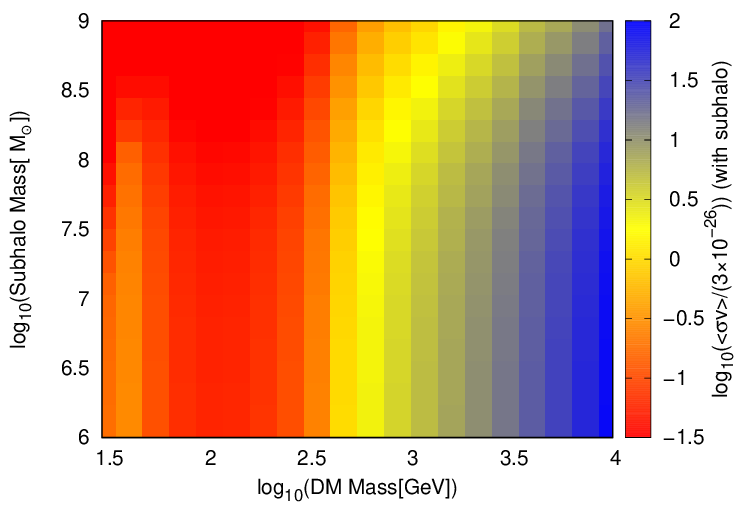}}
\subfigure[]{
  \includegraphics[width=1.0\columnwidth, angle=0]{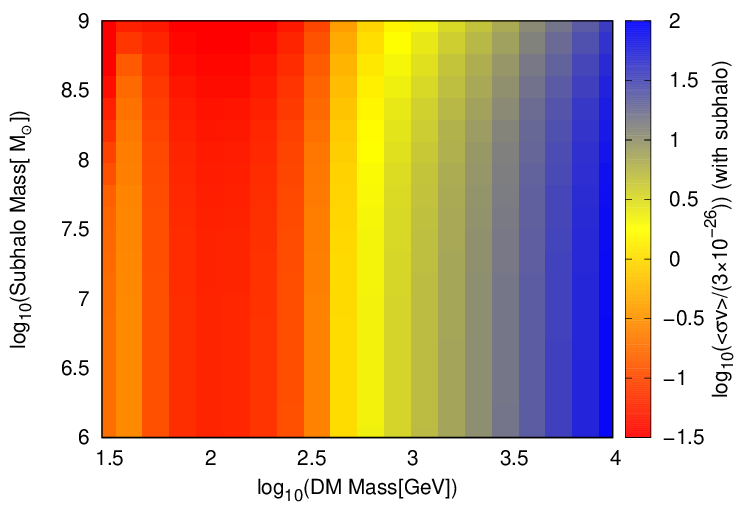}}
\\
\subfigure[]{
  \includegraphics[width=1.0\columnwidth, angle=0]{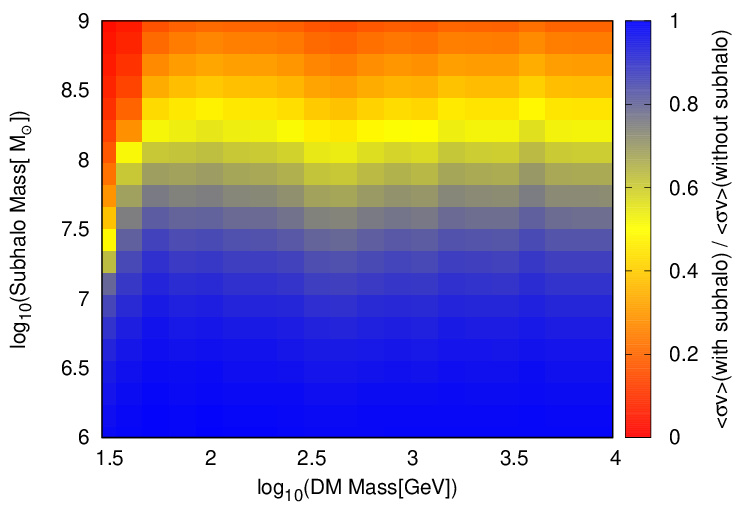}}
\subfigure[]{
  \includegraphics[width=1.0\columnwidth, angle=0]{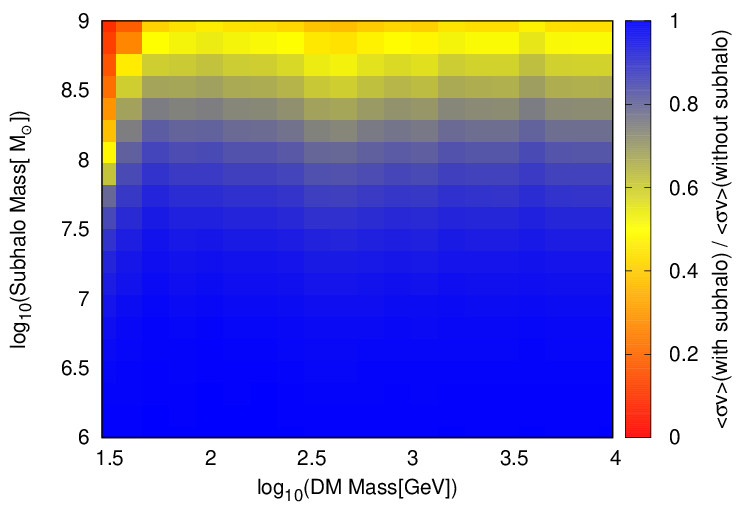}}
\caption{The upper limits on $\langle\sigma v\rangle$ for the $b\bar{b}$
channel at $95\%$ C.L. with the contribution from the nearby subhalo at the distances of 0.3 kpc (a) and 1 kpc (b) from the subhalo are shown in panels (a) and (b), respectively.
The ratios of $\langle\sigma v\rangle$ limits for the scenarios
with and without the nearby subhalo at the distances of 0.3 kpc and 1 kpc are displayed in panels (c) and (d), respectively.}
\label{fig2abcd}
\end{figure*}

In this work, we use the latest AMS-02 antiproton result~\cite{AMS:2021nhj} to set the $95\%$ C.L. upper limits on the DM $\langle\sigma v\rangle$. The mass of nearby subhalo is considered in the range of $10^6-10^9 M_{\odot}$. To enable a comparison, we adopt the same set of propagation parameters in all calculations, but $\Phi$ and $\kappa$ are left to be free. In Fig. \ref{fig2abcd} (a) and (b), we illustrate the resulting limits on $\langle\sigma v\rangle$ for the $b\bar{b}$ channel with the subhalo distances of 0.3 kpc and 1 kpc, respectively. We present the results in the form of a ratio to $\rm 3\times10^{-26}cm^{3}s^{-1}$, which is the so-called natural value corresponding to the correct DM relic density from thermal production. For a certain DM mass, the $\langle\sigma v\rangle$ limits vary indistinctively with the change of subhalo mass.

We present the ratios of the limits on $\langle\sigma v\rangle$ in the scenarios with and without the nearby subhalo at the distances of 0.3 kpc and 1 kpc in Fig. \ref{fig2abcd} (c) and (d), respectively. The existence of a nearby subhalo makes the limits more stringent, resulting in values of the ratio in Fig. \ref{fig2abcd} (c) and (d) being less than 1. We can see that in Fig. \ref{fig2abcd} (c) with the subhalo distance of 0.3 kpc, the limits are minimally influenced by the subhalo masses below approximately $10^6 M_{\odot} \sim 10^7 M_{\odot}$, as depicted by the blue regions. The subhalo masses up to about $10^8 M_{\odot} \sim 10^9 M_{\odot}$ have a significant impact on the limits, as shown by the yellow to red regions. The value of the deepest red region in Fig. \ref{fig2abcd} (c) indicates that the limits on the DM $\langle\sigma v\rangle$ with a subhalo are about one order of magnitude lower than the limits without a subhalo. Compared to the case of 0.3 kpc, a subhalo at a greater distance like 1 kpc leads to a weaker
constraint on the DM $\langle\sigma v\rangle$, and its presence
also has a less pronounced impact on the limits.

\begin{figure*}[!htb]
\centering
\subfigure[]{
\includegraphics[width=1.0\columnwidth, angle=0]{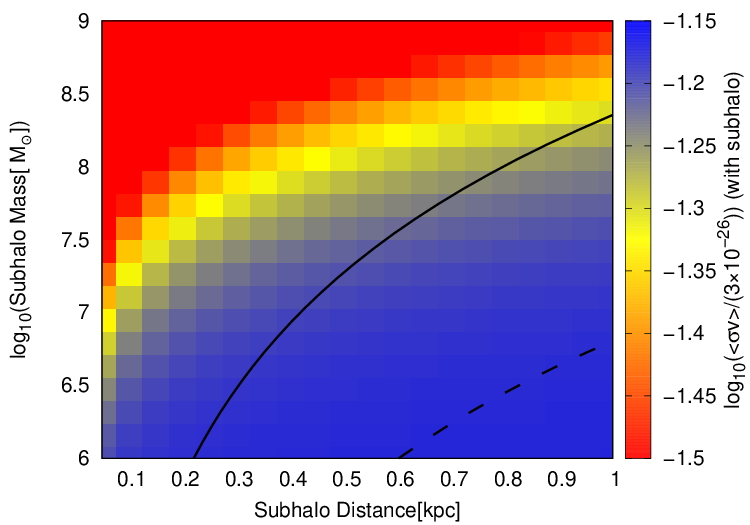}}
\subfigure[]{
\includegraphics[width=1.0\columnwidth, angle=0]{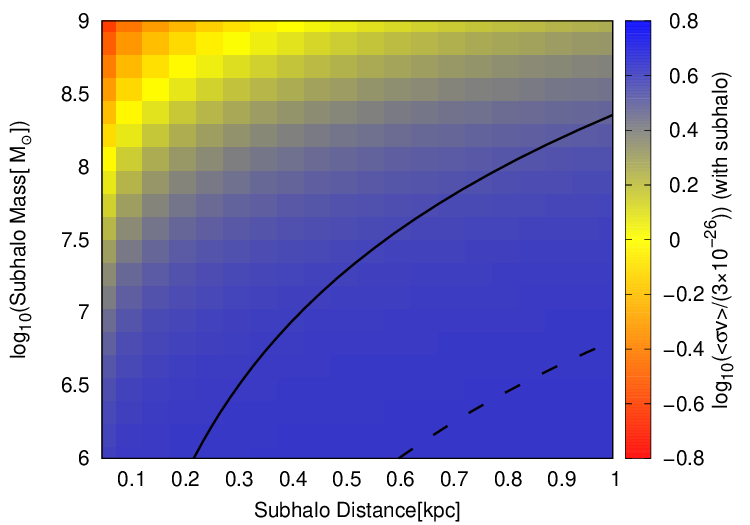}}
\\
\subfigure[]{
\includegraphics[width=1.0\columnwidth, angle=0]{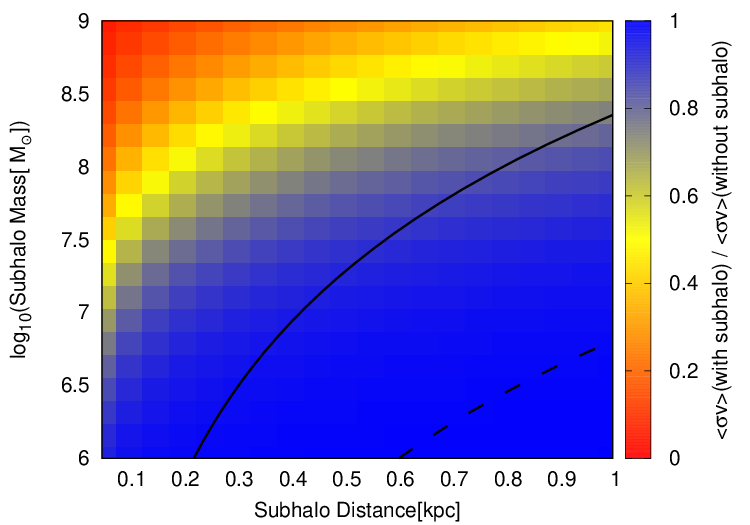}}
\subfigure[]{
\includegraphics[width=1.0\columnwidth, angle=0]{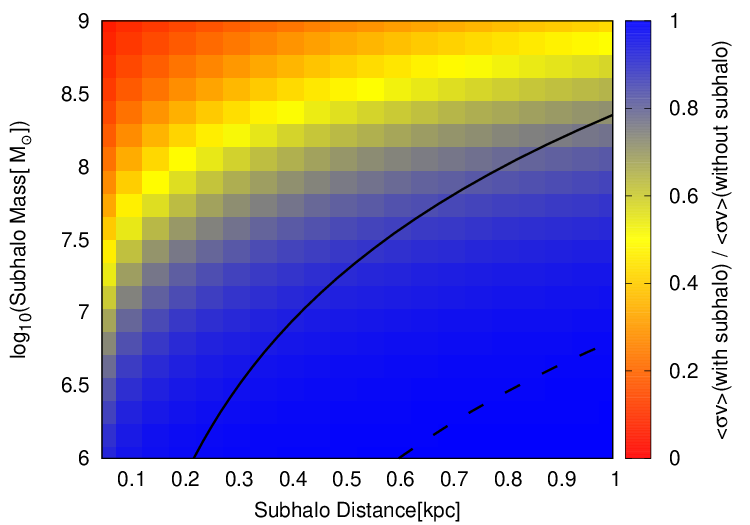}}
\caption{The upper limits on $\langle\sigma v\rangle$ for the $b\bar{b}$
channel at $95\%$ C.L. for the DM masses 75 GeV and 1 TeV are shown in panels (a) and (b),  respectively.
The ratios of $\langle\sigma v\rangle$ limits for the scenarios
with and without the nearby subhalo for the DM masses 75 GeV and 1 TeV are shown in panels (c) and (d), respectively.
The solid and dashed lines represent that the probabilities of finding
a subhalo with the specific mass and distance are $0.1\%$ and
$1\%$, respectively.}
\label{fig3abcd}
\end{figure*}

We also present the limits on $\langle\sigma v\rangle$ for the $b\bar{b}$ channel at $95\%$ C.L., with respect to various subhalo distances and masses. The limits in the form of a ratio to $\rm 3\times10^{-26}cm^{3}s^{-1}$ for two DM masses of 75 GeV and 1 TeV are displayed in Fig. \ref{fig3abcd} (a) and (b), respectively. We can see that for a given distance, the differences in the $\langle \sigma v \rangle$ limits corresponding to different subhalo masses will increase with the increase of DM mass. The ratios of $\langle\sigma v\rangle$ limits for the scenarios with and without the nearby subhalo for DM masses of 75 GeV and 1 TeV are presented in Fig. \ref{fig3abcd} (c) and (d), respectively. As shown in Fig. \ref{fig3abcd} (c) or (d). When the hypothetical subhalo mass is below $\sim 10^{7} M_{\odot}$ in the blue region, the presence or absence of the subhalo has a minor effect on the $\langle\sigma v\rangle$ limits. However, in the red region, the existence of a subhalo would considerably affect the limits, with a maximum impact of about one order of magnitude. The DM mass does not have a significant impact on this result. This can be seen from that the ratio of the $\langle\sigma v\rangle$ limits for the DM mass of 75 GeV is similar to that for 1 TeV in Fig. \ref{fig3abcd} (c) or (d).

The constraints on $\langle\sigma v\rangle$ for the $W^+W^-$ channel are also investigated. We find that for various subhalo distances and masses, the limits are comparable to those of the $b\bar{b}$ annihilation channel with minor differences, due to the similar initial contributions of these two channels.

In Ref.~\cite{Brun:2009aj}, the probability distribution has been provided for finding a particular subhalo with the given annihilation luminosity $\mathcal{L}=\int\rho^2 dV$ and distance,
which is inferred from the N-body simulation Via Lactea II~\cite{Diemand:2008in}.
Using this result, we show the probabilities of $0.1\%$ and $1\%$ for finding a particular subhalo as the black solid and dashed lines in Fig. \ref{fig3abcd}, respectively.

In order to take into account the impact of the concentration model, we utilize an alternative model for comparison
(referred to as model 2) provided by Ref.~\cite{Maccio:2008pcd}
\begin{eqnarray}\label{eq6}
\log C= 0.971 - 0.094 \log \frac{M_{\rm sub}}{10^{12}M_{\odot}}.
\end{eqnarray}

\begin{figure}[!htb]
\centering
\includegraphics[width=1.0\columnwidth, angle=0]{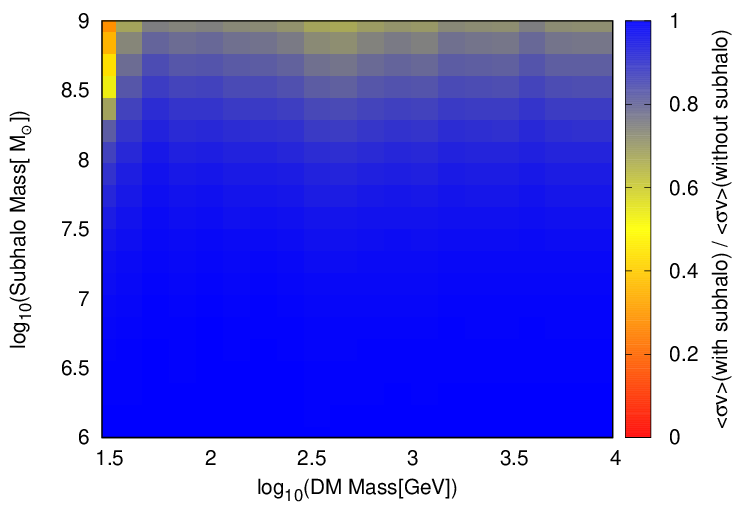}
\caption{The same as Fig. \ref{fig2abcd} (c), except that the concentration model 2 is considered here.}
\label{fig4}
\end{figure}

Taking the subhalo distance as 0.3 kpc, we present the same analysis as Fig. \ref{fig2abcd} (c) in Fig. \ref{fig4}. In comparison with the results in Fig. \ref{fig2abcd} (c), Fig. \ref{fig4} exhibits no red region where the nearby subhalo could significantly affect the constraints. This indicates that the presence of a subhalo
with lower concentration has a relatively smaller impact on the constraints of the DM $\langle\sigma v\rangle$. It is not strange that the concentration parameter has a significant impact on the results, because this parameter determines the shape of the subhalo.
The concentration in model 2 is smaller than that in model 1 for the same subhalo mass.
A smaller concentration implies that the material distribution within this subhalo is relatively scattered,
with a relatively larger core. Such subhalos may have undergone more interactions and merging events
during their formation and evolution, leading to a less centralized distribution of matter. They would provide less contributions to cosmic rays due to the smoother density profiles.

A single nearby subhalo is assumed in the above analysis. The existence of a
wealth of subhalos would further enhance the DM annihilation flux compared with the smooth DM distribution. This enhancement is often described by the boost factor, which
is defined as the ratio of the annihilation luminosity including the subhalo contribution to that originating from the smooth DM component.
The boost factor at a Galactocentric
radius $r$ can be given by
\begin{eqnarray}\label{eqB}
B(r)=\frac{\int\rho^2dV}{\int\bar{\rho}^2dV}=\int_{0}^{\rho_{max}}P(\rho,r)\frac{\rho^2}{\bar{\rho}^2}d\rho,
\end{eqnarray}
where $\bar{\rho}$ is the mean density which is roughly equal to the smooth DM density, and $P(\rho,r)$ is the probability distribution that the particular position at $r$ has the density $\rho$.

With the $P(\rho,r)$ determined by N-body simulations, the boost factor can be further given by ~\cite{Kamionkowski:2010mi,Kamionkowski:2008vw}
\begin{eqnarray}
\label{eqB1}
B(r) = f_se^{\Delta^2} + (1-f_s) \frac{1+\alpha}{1-\alpha}\left[ \left( \frac{\rho_{\rm max}}{\rho_h}\right)^{1-\alpha}-1\right],
\end{eqnarray}
where $f_s\sim 1$ is the fraction of the smooth component in the DM density, and $\rho_h$ is the smooth DM density distribution. The first term resulting from the finite width $\Delta$ of the smooth density is roughly equal to 1 due to small $\Delta$. The second term comes from the contribution of subhalos. The analysis of Ref.~\cite{Kamionkowski:2010mi} shows that $f_{s}(r)$, $\alpha$, and $\rho_{\rm max}$ can be taken as $1-7\times10^{-3}[\bar\rho(r)/\bar\rho(r=100\rm kpc)]^{-0.26}$, 0, and $80\;\GeV\cdot\cm^{-3}$, respectively.

Taking the DM masses of 75 GeV and 1 TeV as examples, we conduct the same
analysis as shown in Fig. \ref{fig3abcd}. Specifically, we derive the upper limits on
the DM $\langle\sigma v\rangle$ in the presence of subhalos, and
the ratio of the upper limits on the DM $\langle\sigma v\rangle$ with and without
subhalos. When investigating the impact of subhalos, both the contributions from the single nearby subhalo and a series of subhalos are considered. The results reveal that
the upper limits on  $\langle\sigma v\rangle$ with the contribution from a series of subhalos is
slightly stricter than that without such contribution by a factor of $\sim 0.8-0.9$.

\section{DM annihilation to $e^\pm$ and $\bar{p}$ with the nearby subhalo}

The aforementioned findings have practical implications for the following discussion. The DAMPE collaboration reported the measurement of the overall spectrum of electrons and positrons ranging from tens of GeV to several TeV, which revealed a tentative peak structure at around 1.4 TeV~\cite{DAMPE:2017fbg}. High energy electrons and positrons at $\sim$TeV cannot travel long distances in the Milky Way due to the strong cooling effect resulting from synchrotron radiation and inverse Compton scattering. A favored hypothesis to explain this sharp spectral structure is a monochromatic injection of $e^\pm$. However, the astrophysical source cannot directly generate such a monochromatic flux, and the required source should be young and close to Earth. On the other hand, DM annihilation in a nearby subhalo may be a promising explanation for this tentative sharp structure ~\cite{Yuan:2017ysv}.

We employ a similar approach to produce the peak feature in the $e^\pm$ spectrum. In order to account for the DM contribution to $e^\pm$, we should consider the rate of energy loss for $e^\pm$ in propagation. This rate is approximated as
\begin{eqnarray}\label{eq5}
  -\frac{\mathrm{d}E}{\mathrm{d}t}\equiv b(E) = b_0 + b_1\frac{E}{1\GeV} + b_2\left( \frac{E}{1\GeV} \right)^2,
\end{eqnarray}
where $b_0\approx 3\times 10^{-16}$\ GeV/s and $b_1\approx 10^{-15}$\ GeV/s represent the rates of energy loss caused by the ionization and bremsstrahlung processes in neutral gas with a density of $1\ \cm^{-3}$, respectively, and $b_2\approx 10^{-16}$\ GeV/s represents the rate of energy loss caused by the synchrotron and inverse Compton scattering processes. The total energy density of the magnetic field and interstellar radiation field is assumed to be 1 $\eV/\cm^{3}$.

We compute the $e^\pm$ spectrum resulting from DM with a mass of 1.5 TeV through the $e^+e^-$ channel for a range of assumed subhalo masses and distances. The subhalo mass and distance are taken from $10^6-10^9 M_{\odot}$ and $0.05-0.5$ kpc, respectively. Then we obtain the best fit $\langle \sigma v \rangle _{e^\pm}$ accounting for the tentative sharp structure in the DAMPE  $e^\pm$ spectrum.

Meanwhile, we utilize the antiproton spectrum observed by AMS-02 to constrain the DM $\langle\sigma v\rangle$ for such nearby subhalo. Although the leptonic channels do not directly produce antiprotons, the gauge bosons originating from annihilation via including the electro-weak correction can decay into antiprotons \cite{Ciafaloni:2010ti,Ciafaloni:2011sa}, despite the low flux. Therefore, there may exist a correlation between the $e^\pm$ and antiproton signals induced by DM even for the leptonic annihilation channels \cite{DeSimone:2013fia}.
We use PPPC~\cite{Cirelli:2010xx,Ciafaloni:2010ti} to calculate this antiproton spectrum from DM annihilation via the $e^+e^-$ channel and derive the corresponding $\langle \sigma v \rangle _{\bar{p}}$ limit at $95\%$ C.L..

\begin{figure*}[!htb]
\centering
\subfigure[]{
\includegraphics[width=1.0\columnwidth, angle=0]{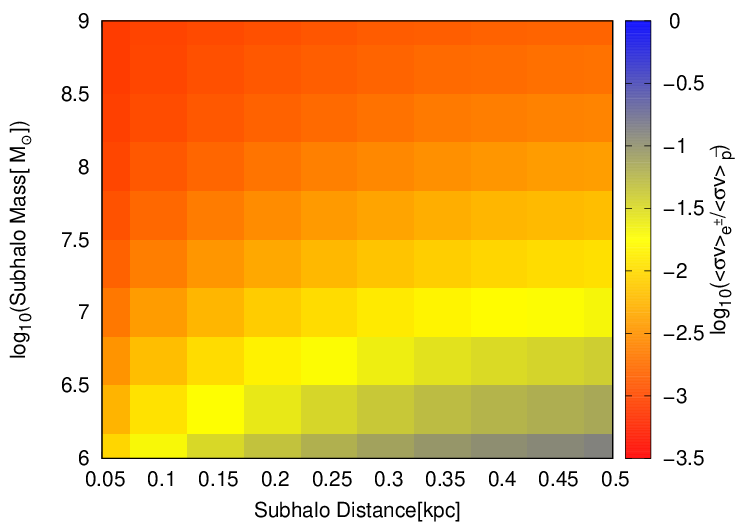}}
\subfigure[]{
\includegraphics[width=1.0\columnwidth, angle=0]{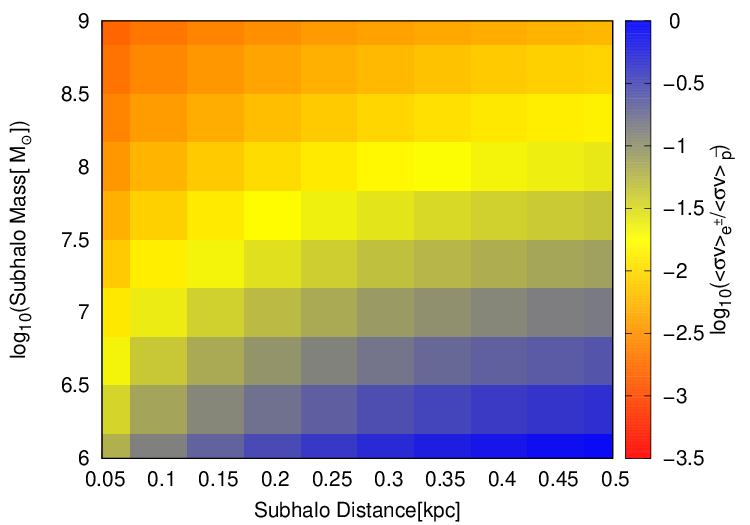}}
\caption{The ratios of $\langle \sigma v \rangle _{e^\pm}$ to $\langle \sigma v \rangle _{\bar{p}}$ for DM with a mass of 1.5 TeV.
$\langle \sigma v \rangle _{e^\pm}$ is the best fit value accounting for the tentative $e^\pm$ spectral  structure at $\sim 1.4$ TeV reported by DAMPE. $\langle \sigma v \rangle _{\bar{p}}$ is the upper limits at $95\%$ derived from the AMS-02 antiproton measurement for the same subhalo parameters.
The left and right panels represent the results for the concentration model 1 and 2, respectively.}
\label{fig5}
\end{figure*}

In Fig. \ref{fig5}, we show the ratios of $\langle \sigma v \rangle _{e^\pm}$ to $\langle \sigma v \rangle _{\bar{p}}$ with the two concentration models, considering different subhalo masses and distances. All the ratios in Fig. \ref{fig5} are smaller than 1. This means that there is no excluded region placed by the  AMS-02 antiproton measurement. From the trends depicted by Fig. \ref{fig5}, we can infer that
the subhalos with smaller masses and concentrations are
more likely to be excluded by the antiproton observation, when explaining the tentative sharp structure in the $e^\pm$ spectrum. This is because that such subhalos have smaller annihilation luminosities and require larger $\langle\sigma v\rangle$ accounting for the DAMPE $e^\pm$ spectrum, which also induces larger antiproton flux.

\section{Conclusion}

In this study, we investigate the potential impact of a nearby DM subhalo on the constraints of DM $\langle \sigma v \rangle$ inferred from the AMS-02 cosmic ray antiproton observation. We separately consider the scenarios of DM annihilation generating antiprotons with and without the nearby subhalo, and present the ratios of the $\langle \sigma v \rangle$ limits for the two scenarios. 
Within the distance of 1 kpc, for the subhalo mass ranging from $10^6 M_{\odot}$ to $10^9 M_{\odot}$ and DM mass ranging from 30 GeV to 10 TeV, the influence of the nearby subhalo on the $\langle \sigma v \rangle$ limits can be at most about one order of magnitude stricter than the case without the subhalo.

We explore the impact of the concentration parameter on the results. 
Two concentration models are employed for comparison. Since the lower concentration implies a more dispersed DM density within the subhalo, the less concentrated model is anticipated to yield a diminished effect on $\langle \sigma v \rangle$ limits.
Due to the potential enhancement of DM annihilation products caused by a wealth of subhalos in the Milky Way, we performe the previous analysis with an additional subhalo distribution. Taking the cases of 75 GeV and 1 TeV DM as examples, our results indicate that the $\langle \sigma v \rangle$ limits with the contribution from the subhalo distribution are slightly stricter than those without such contribution by a factor of $\sim 0.8-0.9$.

Finally, we apply the antiproton constraint on the DM interpretation of the tentative peak structure observed in the DAMPE cosmic ray $e^\pm$ spectrum at $\sim$1.4 TeV. By fitting the DAMPE $e^\pm$ spectrum, we obtain the best fit $\langle \sigma v \rangle _{e^\pm}$ from a hypothetical nearby subhalo for various subhalo masses and distances. Simultaneously, we derive the upper limits from the antiproton measurement for the same subhalo parameters.
We find that the current AMS-02 antiproton results do not set the limit on the DM interpretation accounting for the DAMPE $e^\pm$ spectrum at $\sim$1.4 TeV.

\acknowledgments{This work is supported by the National Natural Science Foundation of China
under Grants Nos. 11947005, 12175248, 12205388, and the scientific research project of Tianjin Education Commission No. 2020KJ003.
}

%\begin{thebibliography}{99}

%\end{thebibliography}

\bibliography{antiproton}

\end{document}